\documentclass{article}

\usepackage{arxiv}
 \usepackage{latexsym,amsfonts}
\usepackage{graphicx}
\usepackage{algorithm}
\usepackage{algorithmic}
\usepackage{multirow}
\usepackage{amssymb}
\usepackage{mathrsfs}
\usepackage{amsmath}
\usepackage{amssymb}%花体字母加粗
\usepackage{mathrsfs}%花体字母
\usepackage{CJK}
\usepackage{multicol} %用于实现在同一页中实现不同的分栏
\usepackage{hyperref}

\newcommand{\beit}{\begin{itemize}}
\newcommand{\eit}{\end{itemize}}

\newsavebox{\savepar}

\def\setsize{\csname @setfontsize\endcsname \setsize}

\begin{document}
\title{Purpose-based Access Policy on Provenance and Data Algebra}
\author{Faen Zhang, Xinyu Fan, Wenfeng Zhou, Pengcheng Zhou\\
	\{faenzhang, pengchengzhou\}ainnovation@gmail.com\\
	fanxinyu@ainnovation.com, zhouwenfeng@interns.ainnovation.com\\
	AInnovation Technology Ltd.} 
\maketitle              % typeset the header of the contribution
\begin{abstract}
It is a crucial mechanism of access control to determine that data can only be accessed for allowed purposes. To achieve this mechanism, we propose purpose-based access policies in this paper. Different from provenance-based policies that determine if a piece of data can be accessed or not, purpose-based access policies determines for what purposes can data be accessed. Particularly, the purposes can be classified as different sensitivity levels. For the first time, We tailor policy algebras to include internal and external policy operators for hierarchical purposes, in order to merge purpose sets generated by individual policies. We also created external policy algebras to merge policies from multi-parties. With different types' testing experiments, our model is proved to be feasible and practical.

%\keywords{First keyword  \and Second keyword \and Another keyword.}
\end{abstract}
\section{Introduction} 
In computer systems, the provenance of a piece of data logs processes and operations which result in the piece of data and is relevant to the source and origins. Provenance can be expressed as a directed acyclic graph (DAG),  illustrating how a data artifact is processed by an execution. In such a DAG of provenance under the Open Provenance Model (OPM)\cite{MoreauCFFGGKMMMPSSB11}, nodes present three main entities including \emph{Artifact}, \emph{Agent} and \emph{Process} and edges represent connections to the main entities. Data provenance logs historical operations performed on documents, preserving its security and privacy. Provenance Access Control is considered a crucial research topic for big data security. The sensitivity of files and their provenance can be different, and users can request, and be granted, access to files and provenance separately. In some situations, provenance itself may consist of sensitive information which might require more protection than its attached document. For instance, although a programming project can be published to the public, its authors and executed operations should be kept as a secret, to prevent leaking the techniques. Therefore, access control to the provenance data itself is required. It allows eligible users to access the provenance data and protects it from unauthorised access.

In provenance-aware systems, traditional access control policies cannot solve all problems, such as protect data security and privacy and ensure proper access. Only decisions are made about whether the pieces of data that are allowed to be accessed cannot meet all the requirements of security protection. Access policies specify laws or preferences for intended purposes, retention, condition, obligation \emph{etc.} They also confine accessibility of data, such as the purposes for which it can be accessed. Hence, purpose-based access policies can specify restrictions that traditional access control policies can not realise, including determining allowed and prohibited access purposes based on provenance. Provenance can contribute to access policies. Note that provenance is closely connected with access policies and can provide crucial information regarding the sensitivity of data because provenance is a file which records historical operations performed on a piece of data, where the history operations can be accessed with the intended data. Hence, provenance can be employed as conditions to map intended usages. For example, an access policy can be defined as: only if a piece of data is submitted to an educational institution, it can be used for the purposes of \emph{research} and \emph{education}. Clearly, access purposes in our provenance-based access policies are determined by whether the provenance contains the given partitions. 

Several existing works proposed access policies \cite{ByunL08}\cite{LinHZ16} which map attributes in queries, roles and system conditions to permitted usages, retention, condition, obligation, \emph{etc.} Byun \emph{et al.} \cite{ByunL08} presented a comprehensive approach for privacy-preserving access control based on the notion of purpose. In their model, purpose information associated with a given data element specifies the intended use of the data element. A key feature of their model is that it allows multiple purposes to be associated with each data element and also supports explicit prohibitions, thus allowing privacy officers to specify that some data should not be used for certain purposes. A discussion arisen by Byun \emph{et al.}\cite{ByunL08} is that it would be more advantageous to organize purposes in a directly acyclic graph instead of a tree construction. 

However, In our framework, an acyclic graph is a better option to organise purposes, because it can support the possibility that a node could have more than one ancestor and where it is closer to the nature of the relationship of purposes. 
Based on their ideas, we utilise provenance subgraphs as conditions for access policies. Because both the previously completed process and the people who execute the processes are critical in determining the purpose for which the data can be accessed. In our proposed provenance-based access policy, we not only employ provenance partitions as conditions, but also distinguish the hierarchies of purposes to understand the different sensitivities of purposes. The hierarchies of purposes occurs mainly when policy results conflict. Another reason to define different hierarchies of purposes is that when merging purpose sets determined by each policy, purposes in different hierarchies should be combined via different operators. For instance, let a collection of purpose be \{\emph{Analysis, Admin, General-Purpose}\}, the sensitivity of the three elements are \emph{Analysis} $>$ \emph{Admin} $>$ \emph{General-Purpose}. It indicates that the \emph{General-Purpose} is the least sensitive purpose, which can be granted to the public, while the purpose of \emph{Analysis} is the most sensitive purpose in the set and can only be accessed by advanced users.

Therefore, policy algebras to merge these purpose sets is tailored. This improves the performance of algebras to distinctively combine purposes in different hierarchies. To the best of our knowledge, previous access policy algebras to merge purpose sets have not distinguished purposes in various hierarchies. Hence, it is attractive to define functions which can combine purposes in different hierarchies by different operators, especially, when combining two collections of purposes. We hope to conjunct usages with a lower rank and to take an intersection of usages with a higher rank. Here, we provide a scenario as the application of this policy. When a request was sent to a database aiming to access a piece of data. The database server refers to the access policies and decides for which purposes can the nominated data being accessed. Let a sample Policy \emph{A} be: if the data was collected by the government, it can be used for data analysis; if the data was reviewed by a university, it can be used for research and education. By confirming with its provenance graph, the data are permitted to access for {data analysis}$_{HH}$ which is the higher hierarchy purpose and {research and education}$_{LH}$ which are the low hierarchy purposes. Similarly, Policy \emph{B} permits the data can be accessed by the same user for {auditing}$_{HH}$ and {education, marketing}$_{LH}$. The database system regulates that higher hierarchy purposes granted by different policies are merged by the operator \emph{union} and lower hierarchy purposes are merged by the operator \emph{intersection}. Hence, the final permitted purposes are \{data analysis, auditing\}$_{HH}$ = \{data analysis\}$_{HH}$ $\cup$ \{auditing\}$_{HH}$, and \{education\}$_{LH}$ = \{research and education\}$_{LH}$ $\cap$ \{education, marketing\}$_{LH}$. 

\subsection{Our Contributions}

In this paper, we propose a framework for purpose-based access policies on provenance. Firstly, we define the semantics and syntax of atomic access policies which map conditions to a set of permitted or prohibited usages. To the best of our knowledge, we are first to utilise provenance as conditions to tailor access policies. We also define how to specify a set of purposes for each policy.
 
Even though purposes were classified based on various sensitivities in previous work, corresponding algebras have not previously been proposed. Concretely, to combine two sets of purposes, purposes with different sensitivities are merged by different operators in this work. Therefore, we particularly design functions for algebras of access policies involving purposes. In addition, the algebras in this framework consist of internal algebra and external algebra. 

This paper was organised as: Provenance-based Access Policies is defined formally in Section 2, where we present its system assumption, syntax, semantics and a case study; In Section 3 and Section 4, we propose the algebras by defining internal and external operators; In Section 5, experiments to evaluate the policies are presented and the conclusion is presented in Section 6.

\section{Purpose-based Access Policy on Provenance}

In this paper, we propose access policies that determine a set of allowed or prohibited purposes based on provenance. As provenance logs operations performed on a piece of data, the access purposes were determined according to the historical records. Historical operations provide crucial clues for the sensitivity of data. Hence, the fine-grained access policies map provenance and attribute in a request to a collection of allowed or prohibited purposes.

\subsection{System Architecture}

\begin{figure}[thb]
\vspace{-0cm}
\centering
\includegraphics[scale=0.35]{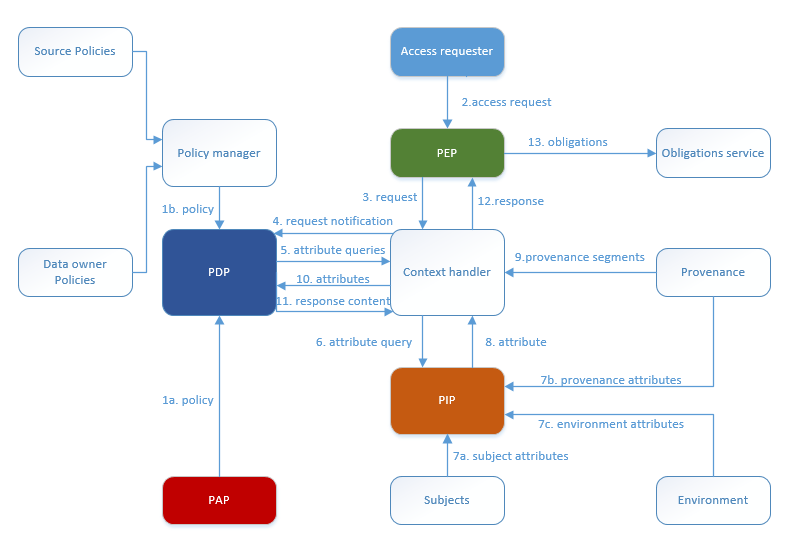}
\vspace{-0.5cm}
\caption{System Architecture} \label{fig1}
\end{figure}

The system architecture for purpose-based access policy is illustrated in Figure 1. The system architecture consists of Policy Administration Point (PAP) which generates access policies and Policy Manager that retrieves policies or results of policies from other parties. Context handler forwards access requests received by Policy Enforcement Point (PEP) to Policy Decision Point (PDP) where the decisions are made according to policies. Moreover, Policy Information Point (PIP) collects required attributes for the policy evaluation from subject, environment, provenance \emph{etc.} Particularly, provenance partitions are retrieved by the context handler and sent to PDP, aiming to generate access purposes. There are two unique features of the system architecture of this framework. One is retrieving provenance partitions as conditions to generate access purposes, the other one is that PDP merges policies from both internal policies and external policies to generate a final set of access purposes for a given piece of data or query. During the process to produce data, it can be executed by more than one data owners who would like to generate their own access policies. It motivates us to propose a policy algebra to merge results of policies from various parties. In the end, the final decision generated by PDP is sent to PEP for enforcement.

\subsection{Semantics} 

Firstly, we define basic elements for the purpose-based access policies, including provenance partitions as atomic policies, access trees to organise atomic policies, purpose graphs, and purpose sets. Purpose graphs which list all possible access purposes are defined by a system, where each policy determines a set of allowed and prohibited purposes in the purpose graphs.\\

\begin{figure}[thb]
\vspace{-0cm}
\centering
\includegraphics[scale=0.35]{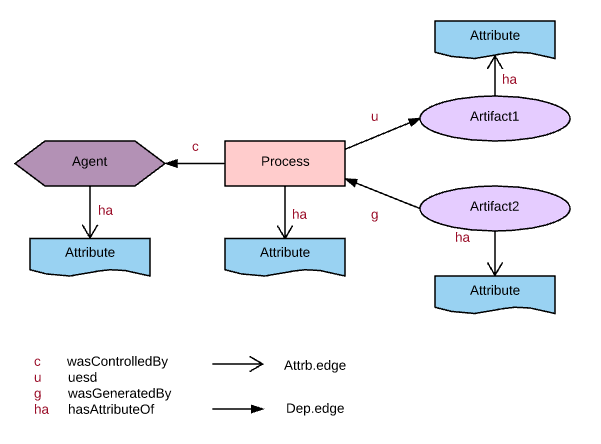}
\vspace{-0.5cm}
\caption{OPM$^+$ Schema} \label{fig1}
\end{figure} 
 
{\bf Definition 1} (Open Provenance Model$^+$(OPM$^+$)) is an extension of OPM, which records how is a piece of data derived. The model is defined by a triple $<$T, L, G$>$:

\begin{itemize}

\item T is the vertex types: agent (Ag), artifact (A), process (P) and attribute (Att). Each vertex in a provenance graph is one of these types. In Figure 2, an artifact is represented by the shape of oval, which is an object or as a piece of data, such as ``$homework_1$", ``\emph{comments}", and \emph{etc.}; a process is an action which is the operation executed on a piece of data, such as ``\emph{submit}" and ``\emph{review}"; an agent is a subject that sponsors an action including ``\emph{$user_1$}" and ``\emph{professor}".   

\item L is the relationship labels: used (u), wasGeneratedBy (wgb), wasControlledBy (wcb), wasTriggeredBy (wtb), wasDerivedFrom(wdf) and hasAttributes (ha).  Each edge in a provenance graph will be labeled as one of these labels. The Labels describe the relationships between the vertices.

\item G is a labelled DAG, where G = $<$V, E$>$, E defines the allowable relationships between the elements, E = \{ (P, A, used), (A, P, wgb), (P, Ag, wcb), (A, A, wdf), (P, P, wtb), (Ag, Att, ha), (P, Att, ha), (A, Att, ha) \}

\end{itemize}

Given the OPM$^+$ $<$T, L, G$>$, an OPM$^+$ instance is defined by a provenance graph $G_i$ = $<V_i, E_i>$, where $V_i$ is a set of entities and $E_i$ $\in V_i \times V_i \times L$. Let $\tau$: $V_i \rightarrow T$ be a function that maps an entity to its type, we say $G_i$ is valid if for each entity v $\in V_i$, $\tau(v) \in$ T, and for each edge (v, v$^\prime$, l) $\in E_i$, ($\tau$ (v), $\tau$ (v$^\prime$), l)$\in$ E. We extend the definition of OPM in paper \cite{ChenENN15}. \\

{\bf Definition 2} (Attribute Node) The extended provenance model attaches context information as attribute nodes in DAG, where each attribute node consists of attribute items and attribute values. For instance, let $Attribute_i$ for a process be \{timestamp: 1/5/2017; system condition: Linux; location: Sydney\}. In this model, provenance can be classified into base provenance data and (optional) attribute provenance data that is associated with main entities in the graph (ag, p, or a). Attribute provenance data is classified into three categories: Agent-related Attributes, Process-related Attributes, and Artifact-related Attributes.  

\begin{itemize}

\item{Agent-related Attributes:} Agents trigger and execute operations, and their attributes include IDs, activated roles \emph{ect.} Identities are usually unique labels to identify users. Activated roles are the roles users employ by taking actions and can play a key role in distinguishing the sensitivity of content and in making access decisions. For instance, when Alice adds a piece of data into a document in the role of teaching assistant, she grades assignments of students. The comments and scores can only be accessed by Alice and the owner of assignments. While when Alice edited data with a role of student, the content of assignments can be read by other students enrolled the same subject. 

\item{Process-related Attributes:} Processes are operations performed on data and result in the change of data, their related attributes include temporal aspects when operations are performed, such as locations, timestamps, system conditions \emph{etc.} They can influence access decisions. For example, operations in provenance can be accessed if they were executed before 2016.  

\item{Artifact-related Attributes:} Artifacts are objects including input messages, output messages, and source data. The related attributes can include object size, permitted usages defined by the data producers \emph{etc.} Some meta-data that might normally be recorded as attributes may not be held in this way in provenance data, as they are recorded within the structure of a provenance graph. For example the generating agent and time of data. 
 
\end {itemize}

Particularly, the dependency linking the attribute node with the main entity is ``att (has attributes of)". There are two approaches to capturing attributes nodes in the graph-based data model \cite{NguyenPS13}. 

\begin{itemize}

\item The attributes of vertices \emph{ag}, \emph{p}, or \emph{a} recorded as individual sets that are attached to the corresponding vertex. 

\item The attributes of vertices \emph{ag}, \emph{p}, or \emph{a} recorded as one set which is attached to the \emph{p} vertex. 

\end{itemize}

Nguyen \emph{et al.} \cite{NguyenPS13} prefer to attach attributes of one operation as one attribute set to \emph{p} vertex.  We choose the former approach, i.e., storing attribute sets separately, where each attribute set links to the node it describes, in order to identify a node directly according to its attached attributes. Specifically, vertices can be defined by their attributes in PACLP. For example, a policy can define all \emph{agent} vertices attached to a role of ``doctor" can be accessed. It can be noticed that the attribute ``doctor" storing at a vertex of \emph{agent} is more convenient to selecting targeted vertices, comparing with storing it at a close \emph{process} vertex. Following, we define partitions of provenance DAG based on the extended OPM model. \\

{\bf Definition 3} (Provenance Partition). We define \emph{provenance partition} which is a connected subgraph in provenance DAG. For a purpose-based access policy on provenance, provenance partitions are utilised as conditions of a policy. Concretely, in our framework, a provenance partition is a collection of vertices in a provenance DAG under the OPM$^+$, which represents one operation or a series of operations recorded in provenance. The historical transactions propose cues for the sensitivity of data and are employed as policy restrictions, which can determine access purposes. Here we provide sample policies to illustrate the connection between provenance and purposes.\\

(1) \emph{If a document has been reviewed by at least three educational experts, then it can be accessed with the aim of education.}

(2) \emph{If a file has been submitted before 2016 and graded by a teaching staff, it can not be accessed for the purpose of revision.}

(3) \emph{If a piece of data was collected by the government, it can not be accessed for academic research.}

(4) \emph{If the data has been uploaded after two different reviews, it can be accessed for market analysis.}

(5) \emph{If the files have been reviewed after 2016, they can not be accessed for audit and direct-use.}

(6) \emph{If the files have been edited by Alice and revised in Sydney, they can be accessed for service-maintenance and service-updates.}\\ 

The conditions of purpose-based access policy on provenance and provenance-based access policy are the same. The difference between them is that the former maps the conditions to access purposes, and the latter maps conditions to a decision value. The decision of access policy indicates if the data can or not be accessed. While access purposes are for which usages can a piece of data to being accessed. Here we define the atomic conditions of purpose-based access policy as:\\

-$null_T$ is a condition;\\
\indent -$(v_{type}, v_{name})$ is a condition, where $v_{type}$ is a vertices type and $v_{name}$ is a vertices name;\\
\indent -$(v_{type}, v_{name}, a, f)$ is a condition, where $v$ is a vertices, $a$ is an attribute value and $f$ is a binary predicate. \\
\indent -($v_{type}, v_{name}$, $x$, $f$) is a condition, where ($v_{type}, v_{name}$) is a vertices, $x$ is an attribute in a query, and $f$ is a binary predicate;\\
\indent -a target is a string of $(v_{type}, v_{name})$ or $(v_{type}, v_{name}, a, f)$ or ($v_{type}, v_{name}$, $x$, $f$) expressed over XPath.\\

A provenance partition is defined over XPath, which is a collection of vertices. It represents historical operations performed on a piece of data. Provenance partitions are conditions of the policies.\\ %Moreover, the attributes of a query could also be utilised as conditions, which is similar to the approach proposed in Chapter 4 of this thesis. We do not repeat it here.\\

{\bf Definition 5} (Atomic Condition Evaluation). The four-valued set for atomic policies is \{$1_p$, $0_p$, $\perp_p$, $\times_p$\}. Between totally matching a provenance partition ($1_p$) and totally not matching it ($\times_p$), there are two values ($0_p$, $\perp_p$) to represent the status between them. Specifically, if a provenance graph meets all attributes of a partition,  it outputs $1_p$ representing a total match; if a provenance graph contains all the types of vertices name in a partitions, but it does not meet all the attribute values, it output $0_p$; if a provenance graph only contains all the vertices types in a partitions, but not meet vertices names and attributes, it outputs $\perp_p$; if a provenance graph does not meet any item, its result is $\times_p$.  \\

\begin{figure}[thb]
\vspace{-0cm}
\centering
\includegraphics[scale=0.35]{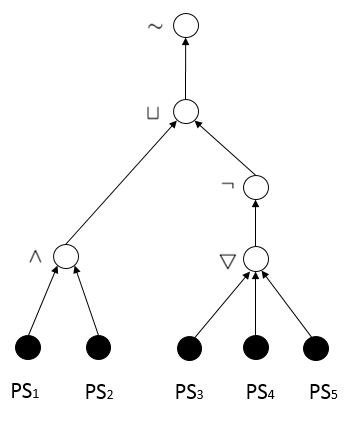}
\vspace{-0.5cm}
\caption{Access Tree of Provenance Segments} \label{fig1}
\end{figure}

{\bf Definition 6} (Purposes). An access purpose \emph{P} implies the aim that a piece of data is being accessed for, such as \emph{military}, \emph{education}, \emph{research}, \emph{etc.} Purpose-based access policies define a collection of allowed and/or prohibited purposes for data. 

In our framework for purpose-based access policies on provenance, purposes with various sensitivities are organised as a form of a directed acyclic graph (DAG). It is denoted as a Purpose Graph (PG). In a PG, each node represents a purpose \emph{P}. Nodes in a \emph{PG} are displayed in several layers and are linked by edges, where the relationship between an ancestor node and a descendant node are generalisation and specialisation. In other words, purposes near the root are more general, while the purposes on the leaf side are more specialised. Hence, each descendent node is a more specialized purpose of its ancestor node. The sensitivity of the hierarchical purposes varies, based on the layers they display. We distinguish the hierarchies of purposes aiming to classify them based on the sensitivities, and further employ functions to merge purposes with different hierarchies by different operations. 

Particularly, there are existing works \cite{ByunL08}\cite{LinHZ16} organising purposes with a tree structure, where each descendant can only have one ancestor. However, in our framework, we replace the tree structure as a DAG. Because in a DAG, a single node can be associated with more than one ancestor. It is more close to the nature of relationships of purposes. We propose an example purpose DAG in Figure 5. In this example, \emph{Admin} is a descendant of \emph{General Purpose}. Hence, it is a more specific purpose comparing with \emph{General Purpose}.\\

\begin{figure}[thb]
\vspace{-0cm}
\centering
\includegraphics[scale=0.45]{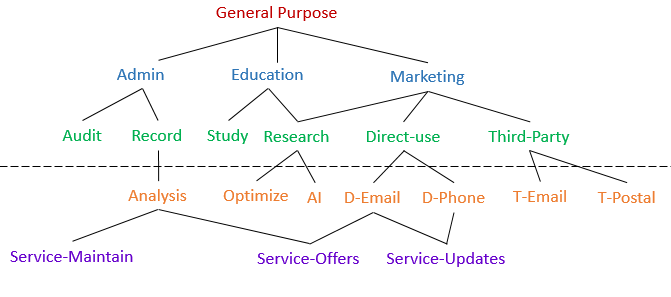}
\vspace{-0.5cm}
\caption{Purpose DAG} \label{fig1}
\end{figure}

{\bf Definition 7} (Purpose Sets). A purpose set defined in each access policy is a collection of purposes in a purpose DAG. Formally, let a set of purposes $\in \emph{PG}$ be a purpose set, which is denoted as $\emph{PS}$.

\begin{itemize}

{\item Ancestors (P)}, denoted by $P^{\uparrow}$. In a \emph{PG}, $P^{\uparrow}$ is the set of all ancestor nodes of \emph{P}, including \emph{p} itself. For instance, in the example purpose DAG provided above, Analysis$P^{\uparrow}$ = \{Analysis, Record, Admin, General Purpose\}. 

{\item Partial Ancestors (P)}, denoted by $P^{\uparrow \alpha}$. In a \emph{PG}, $P^{\uparrow \alpha}$ is the set of $\alpha$ ancestor nodes of \emph{P} including \emph{p} itself. For instance, Analysis$P^{\uparrow 3}$ = \{Analysis, Record, Admin\}.

{\item Descendants (P)}, denoted by $P^{\downarrow}$. In a \emph{PG}, $P^{\downarrow}$ is the set of all ancestor nodes of \emph{P}, including \emph{p} itself. For instance, Admin$P^{\downarrow}$ = \{Admin, Audit, Record, Analysis, Service-Maintain, Service-Offers\}.

{\item Partial Descendants (P)}, denoted by $P^{\downarrow \beta}$. In a \emph{PG}, $P^{\downarrow \beta}$ is the set of $\beta$ ancestor nodes of \emph{P} including \emph{p} itself. For instance, Admin$P^{\downarrow 3}$ = \{Admin, Record, Analysis\}.

{\item $P^{\updownarrow}$} denotes the set of all ancestors and descendants of $P$ in a \emph{PG}, where $P^{\updownarrow}$ = $P^{\uparrow}$ + $P^{\downarrow}$. For instance, Record$P^{\updownarrow}$ = \{General Purpose, Admin, Record, Analysis, Service-Maintain, Service-Offers\}.

\end{itemize}

{\bf Definition 8} (Allowed Purpose and Prohibited Purpose). A purpose set could be denoted as Allowed Purpose (AP) and Prohibitive Purpose (PP). Namely, AP indicates a collection of permissible purposes by which users can access for, while PP indicates a collection of prohibited purposes which not allow access for users. \\

\subsection{Syntax} \label{Syntax}

%Following the semantics of elements of purpose-based policies on provenance, we define its syntax. The syntax is a carrier of policy models. We select a syntax which is different from the XACML proposed in Chapter 3, because the conditions for purpose-based access policies are relatively less. Therefore, we believe the applied syntax is more close to the nature of this policy model. 

Generally, the policy model is tailored under the assumption that an expression of conditions (provenance partitions) maps to a collection of purposes. Namely, if a provenance contains certain provenance partitions, then it can be accessed for a collection of purposes and should be prohibited by another set of purposes. \\

{\bf Definition: Policy Type 1} \{Provenance Partitions$\rightarrow$AP\} In one type of purpose-based access policies on provenance, it maps a tree structure of provenance partitions to a collection of allowed purposes. It indicates that if provenance partitions in a targeted provenance graph meet the access tree structure, the provenance graph can be accessed by the set of allowed purposes. 

\begin{center}
$\mathscr{F}$ ($AP, PG_{i}$, $Access Tree_j$) = \{AP $|$ $\exists Prov_{Partition} \in PG_{i}$ $\&$ $Prov_{Partition} \subseteq Access Tree_j$ \}
\end{center}

The inputs are a targeted provenance graph and an access tree which is the policy. The access tree consists of provenance partitions as leaf nodes and operators as non-leaf nodes. The evaluation result for each provenance partition is value in a four-valued decision set. The operators which are non-leaf nodes merge values for all non-leaf nodes and generate final results for an access tree. If the result is $1_P$, which implies that the provenance matches the access tree. Further, the purposes of the policy can be granted.

{\bf Definition: Policy Type 2} \{Provenance Partitions$\rightarrow$PP\} This maps a combination of provenance partitions to a collection of prohibited purposes, which defines if provenance partitions in a targeted provenance graph meet the access tree structure, then the data can not be accessed for the set of prohibited purposes.

\begin{center}
$\mathscr{F}$ ($PP, PG_{i}$, $Access Tree_j$) = \{PP $|$ $\exists Prov_{Partition} \in PG_{i}$ $\&$ \\
$Prov_{Partition} \subseteq Access Tree_j$ \}
\end{center}

{\bf Definition: Policy Type 3} \{Provenance Partitions $\rightarrow$ AP$|$PP\} This maps a combination of provenance partitions to a collection of allowed purposes and prohibited purposes. It grants a set of allowed purposes and a set of prohibited purposes to the request.

\begin{center}
$\mathscr{F}$ ($AP\&PP, PG_{i}$, $Access Tree_j$) = 

\{AP$\&$PP $|$ $\exists Prov_{Partition} \in PG_{i}$ $\&$ $Prov_{Partition} \subseteq Access Tree_j$ \}
\end{center}
~

{\bf Definition: Policy Type 4} \{Provenance Partitions$|$Data Labels$\rightarrow$AP$\&$PP\} However, in an access policy, the input is not restricted to provenance partitions. While it may include data labels from the data and queries. It includes the subject that is the users attempt to access the data, categories of a piece of data \emph{etc.} For the example below, when the targeted provenance graph meets the access tree, the requestor is within the list of applicable subjects, and the categories of data are within the list of applicable category, a collection of permissible and/or prohibited access purposes can be applied. 

\begin{center}
$\mathscr{F}$ ($AP\&PP, PG_{i}$, $Access Tree_j$) = 

\{AP$\&$PP $|$ $\exists Prov_{Partition} \in PG_{i}$ $\&$ $Prov_{Partition} \subseteq Access Tree_j$ \\
$\& \exists s ( (s \in S_n) \wedge (r \preceq s))$
$\& \exists k ((k \in K_n) \wedge (K_D(i) \subseteq k))$\}
\end{center}
~\\

\subsection{Case Study}

We propose a case study to implement our proposed access policies. Under the scenario illustrated in Figure 4, data is transferred from \emph{System of Records A} to \emph{System of Records B}. The provenance-based access policy determines which purposes the data can be accessed for when the data is delivered to \emph{System of Records B}. For each piece of Data (\emph{i}), there are several attributes attaching to the content of data \emph{q = $Q_D(i)$}, which is shown as:\\

\centerline {Data (\emph{i}) = \{\emph{q = $Q_D (i)$}; \emph{k = $K_D (i)$}; \emph{g = $G_D (i)$}; \emph{p = $P_D (i)$} \} }

\begin{itemize}

\item Category k = $K_D$(i) is a tag attached to a piece of data, which is an attribute used to describe properties of the content \emph{q}. For instance, if a piece of data is labeled a category of ``Medical Records", it might be allowed to be used for \emph{diagnose} purpose. Hence, the category is a condition in the policy to determine possible access purposes.
 
\item Provenance Graph \emph{g = $G_D(i)$} is attached with data to record historical transactions performed on data. In this system, we employ OPM$^+$ which takes the form of a directed acyclic graph (DAG). In this framework, provenance is employed as conditions to determine access purposes.
  
\item Purpose \emph{p = $P_D(i)$} lists a set of access purposes attached with Data (\emph{i}), which is generated by the data producers. The final allowed purposes in the repository should be determined by both of the attached purposes and policy results. 

\end{itemize}

Each System of Records (SOR) is attached to System of Records Notices (SORN). A piece of data delivers between two SOR. Let one SOR be the source \emph{s}, and the other one is the repository \emph{r}. The data repository has an associated SORN \emph{n = N(r)} which lists rules. A rule maps tags of data including \emph{k, g, p} for permitted purposes. Similarly, the source database defines rules to determine the purposes. For a piece of data \emph{i}, the intersection of purposes from source, repository and \emph{p = $P_D(i)$} is the permitted usages.\\
\begin{figure}
\vspace{-0cm}
\centering
\includegraphics[scale=0.35]{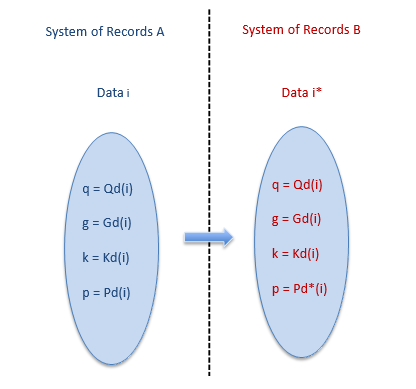}
\vspace{-0.5cm}
\caption{A Case Study} \label{fig1}
\end{figure}

\begin{table*}[htbp]
\centering
\begin{tabular}{lccc}
Data(\emph{i}) = \{$Q_D(i)$; $K_D(i)$=assignment; $G_D(i)$; $P_D(i)$=\{education\} \}\\
\hline
Request: a user as a student asks to access Data(\emph{i})\\
\hline
$<$ \emph{Sample Policy from the Source}$>$\\
\hline
$<Subject>$ students; teaching staff $</Subject>$\\
\hline
$<Category>$ assignments; exam paper $</Category>$\\
\hline
$<Provenance Partitions>$\\
\hline
directed provenance path= ($wasSubmittedBy|Submit, \backslash v^*, wasGradedby|Grade$) \\
\hline
$</Provenance Partitions>$\\
\hline
$<AP>$ education; research $</AP>$\\
\hline
$<PP>$ access investigation $</PP>$\\
\hline
$<$/\emph{Sample Policy from the Source}$>$\\
\hline
$<$ \emph{Sample Policy from the Repository}$>$\\
\hline
$<subject>$ students; teaching staff $</subject>$\\
\hline
$<category>$ assignments; exam paper $</category>$\\
\hline
$<Provenance Partitions>$\\
\hline
directed provenance path= \\($wasSubmittedBy|Submit, \backslash v^*, wasGradedby|Grade$) \\
\hline
$</Provenance Partitions>$\\
\hline
$</Access Tree_{PS}>$\\
\hline
$<AP>$ education; analysis $</AP>$\\
\hline
$<PP>$ research $</PP>$\\
\hline
$<$/\emph{Sample Policy from the Repository}$>$\\

\end{tabular}

\end{table*}

We propose a simple example to illustrate the purpose-based access policies on provenance. Assume a user with a role of \emph{student} asks to access Data(\emph{i}). The policy from the Source determines a set of allowed and prohibited purposes. Under this scenario, the users are allowed to access Data(\emph{i}) for the purposes of \emph{education $\&$ research}. Similarly, the Repository also decides that Data(\emph{i}) can be accessed for \emph{education $\&$ analysis} but not \emph{research}. Finally, the intersection for purposes determined by Source, Repository and $P_D(i)$ is \emph{education}. Hence, Data(\emph{i}) can only be accessed for \emph{education} for the request.

\section{Internal Policies Algebra}

For purpose-based access policies in provenance-aware systems, \emph{internal policy algebras} are required to combine purpose sets generated by individual policies within a system. While \emph{external policy algebras} integrate results of policies from different parties. Firstly, we explore the demands for internal policy algebras.

Under the system assumption of this framework, the sensitivity levels of purposes can be identifiable, because the purposes in policies are defined for the shared purpose DAGs in a system, where the hierarchies of each purpose are initially defined. Hence, it is possible to identify the hierarchy of each purpose within a system and further distinctively process purposes in different hierarchies. Therefore, for internal algebras, we create functions to process purposes with various distinct hierarchies. 

In terms of the motivation, for purposes in DAG are tailored hierarchically, purposes are usually displayed from generalisation to specialisation in the graphs, based on various sensitivities. Semantically, under most scenarios, general purposes are less sensitive and are more easily to grant the permission. While specific purposes are more sensitive and should be cautious to grant the permission. Consequently, purposes in different hierarchies are merged with different operators. For example, it enforces maximum privilege (by \emph{Addition} (+)) for general usage, and minimum privilege (by \emph{Conjunction} ($\&$)) for more specific usages. 

Therefore, in this section, we firstly propose basic operations to merge purposes sets. Then, functions that are constructed by basic operations are defined, in order to propose an approach to merge purposes in distinctly different hierarchies.

\subsection{Basic Operators}

Basic operators are elements of constructed functions for policy algebras, and they are classified as \emph{symmetric operators} and \emph{asymmetric operators}. Specifically, for a symmetric operator, when the two expressions besides it switch, the result does not change. Namely, let $\ast$ be an asymmetric operator, $S_1$ and $S_2$ be two sets of purposes. $S_1 \ast S_2$ = $S_2 \ast S_1$, while, when the sequence of two expressions beside a symmetric operator switch, the result changes. Hence, let $\bullet$ be a symmetric operator, $S_1 \ast S_2 \neq S_2 \ast S_1$. Below, we define 5 asymmetric operators and 1 symmetric operators, which are tailored to merge purpose sets.\\

\begin{table*}
\centering
\begin{tabular}{c c c}
\hline
Operator & Semantics $[\![  ]\!]_e$ & symmetry\\
\hline
\hline
$S_1 + S_2$ & $[\![ S_1 ]\!]_e$ $\cup$ $[\![ S_2 ]\!]_e$ 
& $S_1 + S_2$ = $S_1 + S_2$ (symmetric)\\
$S_1 \& S_2$ & $[\![ S_1 ]\!]_e$ $\cap$ $[\![ S_2 ]\!]_e$ 
& $S_1 \& S_2$ = $S_1 \& S_2$ (symmetric)\\
\hline
$S_1 \boxminus S_2$ & $[\![$ $S_1+S_2$ $]\!]$ - $[\![$ $S_1 \& S_2$ $]\!]$ & $S_1 \boxminus S_2$ = $S_1 \boxminus S_2$ (symmetric)\\
\hline
$S_1 \uparrow \bigtriangleup S_2$ & $S_i$\{I=maximum(Max($S_1$),Max($S_2$))\} 
&$S_1 \uparrow \bigtriangleup S_2$ = $S_1 \uparrow \bigtriangleup S_2$ (symmetric)\\
\hline
$S_1 \downarrow \bigtriangleup S_2$ & $S_i$\{I=minimum(Max($S_1$),Max($S_2$))\} 
&$S_1 \downarrow \bigtriangleup S_2$ = $S_1 \downarrow \bigtriangleup S_2$ (symmetric)\\
\hline
$S_1 \uparrow \bigtriangledown S_2$ & $S_i$\{I=maximum(Min($S_1$),Min($S_2$))\} 
&$S_1 \uparrow \bigtriangledown S_2$ = $S_1 \uparrow \bigtriangledown S_2$ (symmetric)\\
\hline
$S_1 \downarrow \bigtriangledown S_2$ & $S_i$\{I=minimum(Min($S_1$),Min($S_2$))\} 
&$S_1 \downarrow \bigtriangledown S_2$ = $S_1 \downarrow \bigtriangledown S_2$ (symmetric)\\
\hline
$S_1 - S_2$ & $S_1$ - $[\![$ $S_1 \& S_2$ $]\!]$
&$S_1 - S_2 \neq S_1 + S_2$ (asymmetric)\\
\hline
\end{tabular}
\caption{Schema of Basic Operators}
\end{table*}

\begin{itemize}
\item {Union (+)}. Addition of the sets $S_1$ and $S_2$ results in a combined set $S_I$, in which it concludes all the data items from both of the sets. It merges two sets of purposes by returning their union and returns the maximum scope of the purposes of both. This is shown as: 

\begin{center}
$S_I$ = $[\![$ $S_1+S_2$ $]\!]$ = $[\![$ $S_1$ $]\!]$ $\cup$ $[\![$ $S_2$ $]\!]$
\end{center}

\item {Intersection ($\&$)}. Intersection of the sets $S_1$ and $S_2$ returns a set of data items, in which it contains the data items existing at both sets. Namely, it emerges two sets by returning their intersection. Intuitively, intersection returns purposes that in both purpose sets. This can be shown as:

\begin{center}
$S_I$ = $[\![$ $S_1 \& S_2$ $]\!]$ = $[\![$ $S_1$ $]\!]$ $\cap$ $[\![$ $S_2$ $]\!]$
\end{center}

\item {Difference ($\boxminus$)}. The difference of sets $S_1$ and $S_2$ returns data items which only appear in one set but not both. Namely, it returns a set of purposes which are agreed by one policy but not agreed by both.

\begin{center}
$S_I$ = $[\![$ $S_1 \boxminus S_2$ $]\!]$ = $[\![$ $S_1+S_2$ $]\!]$ - $[\![$ $S_1 \& S_2$ $]\!]$
\end{center}

\item {Precedence ($\uparrow$ $\bigtriangleup$)}. It compares the hierarchies of data items with highest hierarchies (the top purposes) in both sets and outputs only one set with the higher top purpose. Particularly, we define two functions to return the highest hierarchy of purposes in the set as Max ($S_m$, $S_n$). Similarly, Min ($S_m$, $S_n$) returns the lowest hierarchy of purposes in the set.

\begin{center}
$S_I$ = $[\![$ $S_1 \uparrow \bigtriangleup S_2$ $]\!]$ (I $\in$\{1,2\}; I = maximum \{Max($S_i$), Max($S_i$)\})
\end{center}

\item {Precedence ($\downarrow$ $\bigtriangleup$)}. This compares the hierarchies of data items with the highest hierarchies in both sets and outputs only one set with the higher top purpose. 

\begin{center}
$S_I$ = $[\![$ $S_1 \downarrow \bigtriangleup S_2$ $]\!]$ (I $\in$\{1,2\}; I = minimum \{Max($S_i$), Max($S_i$)\})
\end{center}

\item {Precedence ($\uparrow$ $\bigtriangledown$)}. This compares the hierarchies of data items with the lowest hierarchies (the bottom purpose) in both sets, and outputs only one set with the higher bottom purpose. 

\begin{center}
$S_I$ = $[\![$ $S_1 \uparrow$ $\bigtriangledown S_2$ $]\!]$ (I $\in$\{1,2\}; I = maximum \{Min($S_i$), Min($S_i$)\})
\end{center}

\item {Precedence ($\downarrow$ $\bigtriangledown$)}. It compares the hierarchies of data items with the lowest hierarchies in both sets, and outputs only one set with the lower bottom purpose. 

\begin{center}
$S_I$ = $[\![$ $S_1 \downarrow$ $\bigtriangledown S_2$ $]\!]$ (I $\in$\{1,2\}; I = minimum \{Min($S_i$), Min($S_i$)\})
\end{center}

\item {Subtraction (-)}. Subtraction of the sets $S_1$ and $S_2$ returns the items in $S_1$ subtracting those appear in $S_2$, which is an asymmetric operator. Namely, when two sets shift positions beside the operator of \emph{subtraction}, the result changes. 

\begin{center}
$S_I$ = $[\![$ $S_1 - S_2$ $]\!]$ = $S_1$ - $[\![$ $S_1 \& S_2$ $]\!]$
\end{center}

\begin{center}
$S_I$ = $[\![$ $S_2 - S_1$ $]\!]$ = $S_2$ - $[\![$ $S_1 \& S_2$ $]\!]$
\end{center}

\end{itemize}

\subsection{Functions for Internal Policy Algebras}

Based on the basic data algebra operators, we define functions to process hierarchical purpose sets. In terms of the motivation, we assume the sensitivity of purposes under a DAG structure is hierarchical. Specifically, the more specific usages should be regarded as the more sensitive ones to be accessed. Therefore, data algebras tend to combine two sets of hierarchical purposes by various operators. Specifically, when combining two sets of hierarchical purposes, we would like to conjunct lower-hierarchy purposes and take the intersection of higher-hierarchy purposes to generate a combined set. \\

Firstly of all, we propose two approaches for dividing a set of purposes as a higher hierarchical subset and a lower hierarchical subset. One approach is that each hierarchy of purpose is defined for the purpose DAG.  Namely, it defines which collection of purposes within a system is higher hierarchically and which collection of purposes is lower hierarchically. Two sample purpose sets from the graph in Figure 3 are presented below to illustrate this. High hierarchical (HH) purposes and low hierarchical (LH) purposes are defined along with the generation of purpose DAGs, where the elements above the dashed line are lower hierarchy purposes and purposes below the dashed line are higher hierarchy purposes. The two sample purpose sets are:\\

$Record^{\updownarrow^1_2}$= \{Admin, Record, Analysis, Service-Maintain, Service-offers\}= \{Admin, Record $\}_{HH}$ + \{Analysis, Service-Maintain, Service-offers$\}_{LH}$

$Marketing^{\updownarrow^1_4}$= \{General Purpose, Marketing, Direct-use, D-Email, D-Phone, Service-offers, Service-Updates\}= \{Admin, Record$\}_{HH}$ + \{Analysis, Service-Maintain, Service-offers$\}_{LH}$\\

Under the default definition of the Purpose DAG, the purposes of the high hierarchy in $Record^{\updownarrow^1_2}$ is \{Admin, Record\}, while the low hierarchy purposes are \{Analysis, Service-Maintain, Service-offers\}. Similarly, the high and low hierarchical sets in $Marketing^{\updownarrow^1_4}$ are \{Admin, Record\} and \{Analysis, Service-Maintain, Service-offers\} respectively. \\

For the other approach to dividing purpose sets, it is determined by the \emph{central purposes} of the sets. The \emph{central purpose} is the keyword to define a set of purposes. In the following examples. \emph{Record} and \emph{Education} are the central purposes. Specifically, when two purpose sets are merged, the central purposes are compared. The central purpose of higher hierarchy determines the position of the parting line, where the row it stays at and below are the high hierarchical purposes. The rest are the low hierarchical purposes. To illustrate this, we still pick up two sample purpose sets from the graph in Figure 3.\\

$Record^{\updownarrow^1_2}$ = \{Analysis, Record, Admin, General Purpose\} = \{Admin, Record $\}_{HH}$ + \{Analysis, Service-Maintain, Service-offers$\}_{LH}$

$Education^{\updownarrow}$ = \{Optimise, AI, Research, Study, Eduction, General Purpose\} = \{Optimise, AI, Research $\}_{HH}$ + \{Study, Eduction, General Purpose$\}_{LH}$\\

To classify each purpose set as a high hierarchical set and a low hierarchical set, the central purposes which are \emph{Record} and \emph{Education} should be compared first. The purposes are displayed from top to bottom according to the order as from generalised purposes to specific purposes, so Hierarchy (Record) $>$ Hierarchy (Education). The row where \emph{Record} stays at and the those below are high hierarchical purposes, and the rest are low hierarchical purposes. Therefore, the high hierarchical collections for each set are \{Admin, Record $\}_{HH}$ and \{Optimise, AI, Research $\}_{HH}$. All the rest are purposes of the low hierarchy.\\

After dividing purposes of each set as two collections based on their hierarchies, we define the functions to merge purpose sets. Let $HA_i$ represents high hierarchy allowed purposes, $HP_i$ represents high hierarchy prohibited purposes, $LA_i$ represents low allowed purposes, and $LP_i$ represents low prohibited purposes, the u-ary operators are defined as below:

\begin{center}
$f_\oplus (S_i, S_j)$ = $S_i \oplus S_j$ = $\frac{(HA_i \& HA_j) - (HP_i - HP_j)}{(LA_i + LA_j) - (LP_i - LP_j)}$
\end{center}

\begin{center}
$f_\ominus (S_i, S_j)$ = $S_i \ominus S_j$ = $\frac{(HA_i \& HA_j) - (HP_i \& HP_j)}{(LA_i + LA_j) - (LP_i \& LP_j)}$
\end{center}

\begin{center}
$f_\otimes (S_i, S_j)$ = $S_i \otimes S_j$ = $\frac{(HA_i \& HA_j) - (HP_i - HP_j)}{(LA_i + LA_j) - (LP_i \& LP_j)}$
\end{center}

\begin{center}
$f_\oslash (S_i, S_j)$ = $S_i \oslash S_j$ = $\frac{(HA_i \& HA_j) - (HP_i \& HP_j)}{(LA_i + LA_j) - (LP_i - LP_j)}$
\end{center}

\begin{center}
$f_\odot (S_i, S_j)$ = $S_i \odot S_j$ = $\frac{(HA_i + HA_j) - (HP_i - HP_j)}{(LA_i + LA_j) - (LP_i \& LP_j)}$
\end{center}

\begin{center}
$f_\uplus (S_i, S_j)$ = $S_i \uplus S_j$ = $\frac{(HA_i + HA_j) - (HP_i - HP_j)}{(LA_i + LA_j) - (LP_i \& LP_j)}$
\end{center}

\begin{center}
$f_\dotplus (S_i, S_j)$ = $S_i \dotplus S_j$ = $\frac{(HA_i + HA_j) - (HP_i \& HP_j)}{(LA_i + LA_j) - (LP_i \& LP_j)}$
\end{center}

\begin{center}
$f_\Cap (S_i, S_j)$ = $S_i \Cap S_j$ = $\frac{(HA_i + HA_j) - (HP_i \& HP_j)}{(LA_i \& LA_j) - (LP_i \& LP_j)}$
\end{center}

\begin{center}
$f_\Cup (S_i, S_j)$ = $S_i \Cup S_j$ = $\frac{(HA_i + HA_j) - (HP_i \& HP_j)}{(LA_i \& LA_j) - (LP_i - LP_j)}$
\end{center}

\begin{center}
$f_\boxtimes (S_i, S_j)$ = $S_i \boxtimes S_j$ = $\frac{(HA_i \boxminus HA_j) - (HP_i - HP_j)}{(LA_i \boxminus LA_j) - (LP_i \& LP_j)}$
\end{center}

\begin{center}
$f_\boxdot (S_i, S_j)$ = $S_i \boxdot S_j$ = $\frac{(HA_i \boxminus HA_j) - (HP_i - HP_i)}{(LA_i + LA_j) - (LP_i \& LP_j)}$
\end{center}

\begin{center}
$f_\boxplus (S_i, S_j)$ = $S_i \boxplus S_j$ = $\frac{(HA_i + HA_j) - (HP_i - HP_j)}{(LA_i \boxminus LA_j) - (LP_i \& LP_j)}$
\end{center}

\begin{center}
$f_\divideontimes (S_i, S_j)$ = $S_i \divideontimes S_j$ = $\frac{(HA_i \boxminus HA_j) - (HP_i - HP_j)}{(LA_i \& LA_j) - (LP_i \& LP_j)}$
\end{center}

The functions merge purposes in different hierarchies with different operators. These functions are utilised for internal algebras of this framework. In terms of which functions should be used and how to combine functions as an expression, this should be determined by each system's default settings or the policies. The integration of policies may involve multiple operators, hence we define the concept of FIDA expressions. \\

{\bf Definition 9} A FIDA expression is defined as:

-If \emph{S} is a set, then \emph{S} is a FIDA expression;

-If $S_1$ and $S_2$ are FIDA expressions, so is a function $f$ ($S_i$, $S_j$).

-If function $f_\alpha (S_i, S_j)$ is a FIDA expression, $f_\beta (S_m, S_n)$ is another FIDA expression, so are $f_\alpha (S_i, S_j) \bullet f_\beta (S_m, S_n)$, where $\bullet$ is a function.\\

In an expression, \emph{Intersection} and \emph{Precedence} operators take high priorities, while Addition and Subtraction take low priorities. For the same priority operators, calculations should be run from left to right. For example, $S_1 \& S_2 + S_3 \triangleright S_4$ is interpreted as $(S_1 \& S_2) + (S_3 \triangleright S_4)$.

A function could also merge more than two purpose sets. A FIDA expression can be defined as in the example below:\\

\begin{center}
$f (S_1, S_2,... S_n)$ = $\frac{\sum_{i=1}^n HA_i - \sum_{i=1}^n HP_i}{\boxminus_{i=1}^n LA_i - \&_{i=1}^n LP_i}$
\end{center}

\section{External Policy Algebra}

External policy algebras are tailored to merge policy results from different parties. Let us imagine the situation where a piece of data was delivered from a source to a recipient. The source, recipient and even data owners generate access policies for the piece of data, after which we propose external policy algebras to merge results from various parties, in order to generate a final intended purpose for the data.

The algebras can be generated by a server viewed as a third party or by the recipient. However, as policies from different authorities cannot be generated from uninformed purpose DAGs, the hierarchies of purposes are not based on the same criteria anymore. Hence, we do not distinguish the hierarchies of purposes when sets of purposes are merged. 

Following, based on basic operators defined in the previous section, we define functions for external policy algebra, where \emph{m} and \emph{n} are two parties. The functions are defined to merge results from two parties. Further, an expression consisting of several functions could merge various parties flexibly. The functions to merge purpose sets from multi-parties are defined below, where \emph{AP} represents allowed purposes and \emph{PP} represent prohibited purposes. $AP_m$ is the allowed purposes for $S_m$; $PP_m$ is the prohibited purposes for $S_m$; $AP_n$ is the allowed purposes for $S_n$; $PP_n$ is the prohibited purposes for $S_n$.    
                                                                                                                                                                                                                                                                                                                                                                                                                                                                                                                                                                                                         
\begin{center}
$F_1 (S_m, S_n)$ = $(AP_m+AP_n) - (PP_m \& PP_n)$

$F_2 (S_m, S_n)$ = $(AP_m+AP_n) - (PP_m - PP_n)$

$F_3 (S_m, S_n)$ = $(AP_m\& AP_n) - (PP_m \& PP_n)$

$F_4 (S_m, S_n)$ = $(AP_m\& AP_n) - (PP_m - PP_n)$

$F_5 (S_m, S_n)$ = $(AP_m \boxminus AP_n) - (PP_m \triangleleft PP_n)$

$F_6 (S_m, S_n)$ = $(AP_m \boxminus AP_n) - (PP_m \triangleright PP_n)$

$F_7 (S_m, S_n)$ = $(AP_m \bigtriangleup AP_n) - (PP_m \boxminus PP_n)$

$F_8 (S_m, S_n)$ = $(AP_m \bigtriangledown AP_n) - (PP_m \& PP_n)$\\
\end{center}

The functions for external algebras generally produce intended purposes as \emph{IP = AP - PP}. However, the basic operators to merge the allowed purposes \emph{AP} and prohibited purposes \emph{PP} are different. Such as $F_1$ combines allowed purposes by \emph{Addition} (+), while it takes the intersection ($\&$) for prohibited purposes.

Moreover, an expression to merge purpose sets from more than two parties could consist of a set of external policy functions. This expression is defined as:\\

{\bf Definition 10} A FIDA expression for the functions of external algebra:

-If \emph{S} is a set, then \emph{S} is a FIDA expression;

-If $S_1$ and $S_2$ are FIDA expressions, so is a function $F$ ($S_i$, $S_j$).

-If function $F_\alpha (S_i, S_j)$ is a FIDA expression, $F_\beta (S_m, S_n)$ is another FIDA expression, so are $F_\alpha (S_i, S_j) \bullet F_\beta (S_m, S_n)$, where $\bullet$ is a function.

\section{Evaluation}

We did experiments to evaluate the performance of the proposed policies and data algebras. We implement the process of policy generation and internal policy algebras and external algebras. The purpose hierarchy has been taken into consideration for the implementation. 

We process the experiments on a 3.40 GHz Intel Computer with 16GB of memory. The machine's operating system is Microsoft Windows 8 and the database is Oracle Database. We set up synthetic datasets based on the version proposed by Wisconsin Benchmark \cite{BittonDT83}. To be more specific, every database consists of 3 numeric and 6 string columns. We set up data values refer to the specification from the paper \cite{BittonDT83}. The provenance model is OPM$^+$, and 200 purposes were generated. Particularly, in the policy generation phrase, the 200 purposes are randomly selected to attach to policies. 

After establishing provenance graphs and purposes, we tested the time span for the policy generation. Four types of policies are tested, which is defined in Section 2.3. The aim to do this experiment is to compare the time costs for each type of policies. The more complicated of a policy contains more restrictions, the longer time requires generating the policy. To measure the time of generating a policy, it counts from generating random provenance subgraphs, then combing with conditions and access purposes randomly. Thus, the reported time takes the whole policy generating process into account. We repeated the process ten times, and take the average time to draw the figure. Following, we implemented the policy algebras proposed in this paper as well. We compare the time span for internal policy algebras and external policy algebras. The experiment also records the average values of ten times of simulation. Notably, the internal policy algebras take longer time compared with the external policy algebras, because in our framework, only the internal policy algebras distinguish the hierarchies of purposes. Hence, internal policy algebras calling more functions take longer time than external policy algebras in the implementation.   
\begin{figure}
%\vspace{-10pt}
\centering
$\begin{array}{cc}
\includegraphics[width=4cm]{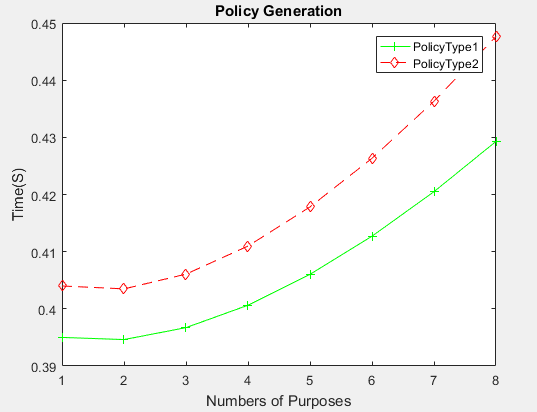} &
\includegraphics[width=4cm]{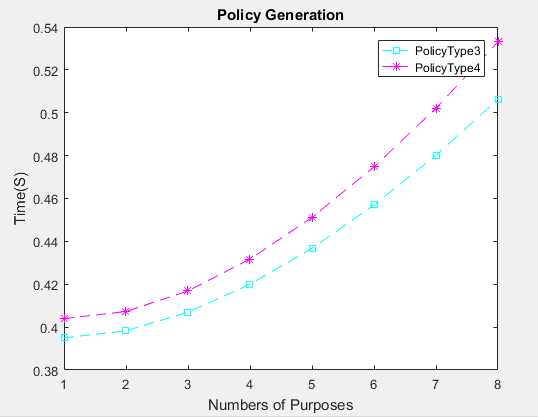} 
\\
%\mbox{(\textsc{lf-lda})} & \mbox{(\textsc{lf-dmm})}
\end{array}$
\caption{Time spans of 4 different basic Policy Type.}
\label{fig:ourlmodels}
\end{figure}

\begin{figure}
\vspace{-0cm}
\centering
\includegraphics[scale=0.3]{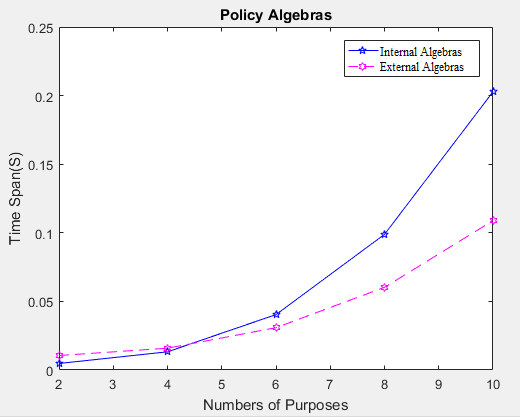}
\vspace{-0.5cm}
\caption{Time spans comparison of internal Algebras and External Algebras.} \label{fig1}
\end{figure}

\section{Conclusion}

In this paper, purpose-based access policies on provenance are proposed, in order to establish a comprehensive scope of access control mechanisms on provenance. First, we define the policy model including syntax and semantics, which maps attributes in provenance to allowed and prohibited purpose sets. The policies determine which purposes a piece of data can be accessed for, and this is based on whether the provenance contains certain provenance partitions. Moreover, as the sensitivities of purposes are different, the purposes are classified as various hierarchies in this framework. 

We define internal and external policy algebras for purpose-based access policies. To the best of our knowledge, it is the first time defining the functions that merge purposes in various hierarchies by different basic operators. Moreover, external policy algebras to merge policies from multi-parties are also created. 

\bibliographystyle{my}
\bibliography{references}

\end{document}